\documentclass[twocolumn,amsmath,amssymb,10pt,aps]{revtex4-1} 

\usepackage{graphicx}
\usepackage{latexsym}
\usepackage{amsmath}
\usepackage{amsthm}
\usepackage{amssymb}
\usepackage{epstopdf} 
\usepackage{enumerate}
\usepackage{setspace} % controllabel line spacing
\usepackage{dcolumn}% Align table columns on decimal point
\usepackage{bm}% bold math
\usepackage{setspace} % controllabel line spacing
\usepackage{slashed}
\usepackage{color}

\begin{document}
\title{
Exotic $Z_2$ Symmetry Breaking Transitions in 2D Correlated Systems 
%\\: Theory of Pseudo-gap Transitions
%\\: Theory of Phase Transitions at Pseudo-gap Temperatures
 }
\author{Sangjin Lee$^1$}
\author{Jun Jung$^1$}
\author{Ara Go$^2$}
\author{Eun-Gook Moon$^1$}
\thanks{egmoon@kaist.ac.kr}
\affiliation{$^1$Department of Physics, KAIST, Daejeon 34141, Korea}
\affiliation{$^2$Center for Theoretical Physics of Complex Systems, IBS, Daejeon 34051, Korea}

\date{\today}

\begin{abstract}  
The Landau paradigm of phase transitions is one of the backbones in critical phenomena. With a $Z_2$ symmetry, it describes the Ising universality class whose central charge is one half ($c=1/2$) in two spatial dimensions (2D). 
Recent experiments in strongly correlated systems, however, suggest intriguing possibilities beyond the Landau paradigm.    
We uncover an exotic universality class of a $Z_2$ symmetry breaking transition with $c=1$. It is shown that fractionalization of discrete symmetry order parameters  may realize the exotic class. In addition to novel critical exponents, we find  that the onset of an order parameter may be { super-linear} in contrast to the sub-linear onset of the Ising class. 
We argue that a super-linear onset of a $Z_2$ order parameter without breaking a bigger symmetry than $Z_2$ is evidence of exotic phenomena, and our results are applied to recent experiments in phase transitions at pseudo-gap temperatures. 
\end{abstract}

\maketitle
Fathoming exotic phenomena beyond the Landau paradigm is one of the most fundamental problems in strongly correlated systems. 
Prime examples include strange metals and pseudo-gap phenomena in high temperature superconductors \cite{Kivelson, Lee, Taillefer, Keimer15, Sachdev, Vojta, erez, Zhang}, and recent advances in topological matter research expand territories of the exotic phenomena  since topology further classifies phases and their transitions \cite{Kitaev, Wen, Senthil, Nakatsuji, Pesin, Moon, Kondo, Herbut, Savary, SachdevTop}. 
To demystify the exotic phenomena, novel concepts such as deconfined phases and phase transitions have been proposed \cite{deconfined, ashvin,SachdevTop}. 

Remarkable developments in recent experiments shed new lights on strongly correlated systems.
New phases with broken discrete symmetries are uncovered especially in cuprates and iridates, and most of them are associated with a $Z_2$ symmetry such as inversion, time-reversal, and nematicity on a square lattice  \cite{Badoux, Wu, Xia, Li, Croft, nem1, nem2, nem3, Lawler, Hsieh1,Hsieh2,bj}. Under the Landau paradigm, their universality class is the celebrated Ising class, but unusual results are reported \cite{Xia, Wu, Badoux,nem1,nem3}. For example, super-linear onsets of order parameters appear at pseudogap temperatures. Motivated by these studies, we investigate the possibility of an exotic $Z_2$ transition in 2D strongly correlated systems.
 
In 2D, conformal field theories (CFTs) characterize continuous phase transitions, and the Landau theory with a $Z_2$ symmetry is  described by the minimal CFT with $c=1/2$, the Ising class \cite{CFT}. 
Then, it is obvious that a non-Ising transition with a $Z_2$ symmetry must be described by a CFT with $c \neq 1/2$. 
This calls for mechanisms beyond the Landau paradigm, and we provide such mechanisms employing fractionalization of discrete symmetry order parameters. 
Striking properties of an exotic $Z_2$ transition with $c=1$ and their applications to pseudogap transitions are discussed below.  

Let us recall the Ising class under the Landau paradigm. A non-trivial representation of a symmetry group is prescirbed to an order parameter. 
To be specific, we consider the representations ($B_1, B_2$) of a square lattice symmetry group, $C_{4v}$, whose  basis functions are $(x^2-y^2, xy)$. Under a reflection, for example $y \rightarrow -y$, ($B_1, B_2$) acts differently. A $90^{\circ}$ rotation ($\mathcal{R}_{{\pi}/{2}}$) acts as a $Z_2$ symmetry in the two representations  because the four fold rotational symmetry is broken down to the two-fold one $(C_4 \rightarrow C_2)$, so-called Ising-nematicity.  
The Landau theory for  $B_1$ becomes
\begin{eqnarray}
S_{Landau} = \int_x \frac{1}{2}( \nabla  B_{1}(x))^2 + \frac{r}{2}  B_{1}(x)^2 +\frac{g}{24}  B_{1}(x)^4 +\cdots, \nonumber
\end{eqnarray}
with coupling constants, $(r,g)$ and a shortened notation $\int_x = \int d^D x$, and including $B_2$  is straightforward. 
In 2D, 
the self-duality of the Ising class is one  of the main characteristics, which can be conveniently understood by the $Z_2$ clock model. 
Using the polar representation of the order parameters, $(B_{1} = \rho \cos(\Phi), B_2=\rho \sin(\Phi))$, the symmetry actions become
\begin{eqnarray}
\Phi &\rightarrow& \Phi+ \pi \quad  (\mathcal{R}_{{\pi}/{2}}), \quad \Phi \rightarrow -\Phi \quad  ({\rm reflection}).   \nonumber 
\end{eqnarray}
Then, the  $Z_2$ clock model is naturally introduced  
$\mathcal{S}_{Z_2} = \int_x \frac{K}{2 \pi} (\nabla \Phi )^2 - u_2 \cos(2 \Phi )$,
with two parameters ($K, u_2$)  in a continuum form.
Its dual theory is   
\begin{eqnarray}
\mathcal{S}_{Z_2, dual}  = \int_x \frac{(\nabla \tilde{\Phi})^2 }{2\pi K} - u_{2v} \cos(2\tilde{\Phi}) -u_2 \cos(2 \Phi),
\end{eqnarray}
with a dual field ($\tilde{\Phi}$) \cite{Jose,dual}. 
The sign of $u_2$ chooses broken-symmetry states,  bond nematicity ($B_1$) for $u_2 >0$ and diagonal nematicity ($B_2$) for $u_2 <0$.
The self-duality is manifested by the same form of the two cosine terms, $(\cos(2\Phi), \cos(2 \tilde{\Phi}))$. 
We stress that the two cosine terms describe different physical quantities in spite of their same form. Namely, the cosine term with $\Phi$ describes energy density of the order parameter, $\cos(2 \Phi) \propto B_{1,2}^2 $, while the cosine term with $\tilde{\Phi}$ describes a topological defect, $2\pi$ vortex, as in the Kosterlitz-Thouless (KT) transition \cite{KT, KT2}. Characteristic interplay between the self-dual cosine terms gives rise to the Ising class.

One obvious way for a non-Ising CFT is to give up the $Z_2$ symmetry adopting a bigger symmetry such as $Z_N$ with $N>2$. For example, vector-type order parameters such as valence bond solid or loop current order break the four-fold rotational symmetry completely ($C_4 \rightarrow C_1$). The $Z_4$ clock model  may describe their phase transitions,
$\mathcal{S}_{Z_4} = \int_x \frac{K}{2 \pi} (\nabla \Phi )^2 - u_4 \cos(4 \Phi )$, where a $\mathcal{R}_{\pi/2}$ rotation acts as a $Z_4$ symmetry $(\Phi \rightarrow \Phi+\pi/2)$, and its dual theory is  
\begin{eqnarray}
\mathcal{S}_{Z_4, dual} = \int_x \frac{(\nabla \tilde{\Phi} )^2}{2 \pi K}  - u_{2v} \cos(2\tilde{\Phi})-u_4\cos(4 \Phi ).
\end{eqnarray}
The different form of the two cosine terms demonstrate the absence of the self-duality, and its criticality is described by the Ashkin-Teller class, a CFT with $c=1$ \cite{ Kadanoff, Ginsparg}.

The above analysis hints that an exotic $Z_2$ transition may be realized by suppressing the $2\pi$ vortex term keeping the $Z_2$ symmetry. 
The suppression is impossible under the Landau paradigm, so mechanisms beyond the Landau paradigm are indispensable. 
Our strategy is to represent order parameters with fractionalized fields. For example, two bosonic fields, $\psi_{1,2} \equiv \sqrt{\rho_{1,2}} e^{i \phi_{1,2}}$, may give physical (gauge-invariant) observables
\begin{eqnarray}
b_1 &=& (\psi^{\dagger}_1 \psi_2 + \psi^{\dagger}_2 \psi_1)/2 = \sqrt{\rho_1 \rho_2} \cos(\phi_1 - \phi_2) , \nonumber \\
b_2 &=& i ( \psi^{\dagger}_1 \psi_2 - \psi^{\dagger}_2 \psi_1)/2 = \sqrt{\rho_1 \rho_2} \sin(\phi_1 - \phi_2). \nonumber 
\end{eqnarray}
An internal $U(1)$ gauge structure, $\phi_{1,2}(x) \rightarrow \phi_{1,2}(x) +\lambda(x) $ with $\lambda (x)\in (0,2\pi)$, is imposed and the symmetry actions are  
\begin{eqnarray}
(\phi_1, \phi_2) \quad &\rightarrow& \quad (\phi_1 + \frac{\pi}{2}, \phi_2 - \frac{\pi}{2}) \quad \quad (\mathcal{R}_{\pi/2}) \nonumber \\
(\phi_1, \phi_2) \quad &\rightarrow& \quad (-\phi_1 , -\phi_2) \quad \quad \quad \quad ({\rm reflection}). \nonumber 
\end{eqnarray} 
Roughly speaking, one may interpret that the angle variable $\Phi$ in the $Z_2$ clock model is split into the two angle variables ($\phi_{1,2}$) with the gauge constraint (see SI). 
We remark that  generalizations to other $Z_2$ symmetries are straightforward. For example, an inversion symmetry with a two fold symmetry group ($C_{2h}$) allows two representations ($A_{u}, B_{u}$) whose fractionalized representations can be similarly obtained.  

Many body physics with the fractionalized discrete order parameters may be investigated by considering a generic lattice Hamiltonian,
\begin{eqnarray}
H &=& \sum_{i,j} t_{\alpha} \big(  \psi^{\dagger}_{\alpha} (i) \psi_\alpha (j) e^{i a_{ij}}  +h.c \big) +\sum_{i_p} \frac{f (i_p)^2}{2 e^2} + \frac{f(i_p)^4}{4 \Lambda} \nonumber \\
&&+ \sum_{i} r_{\alpha} |\psi_{\alpha}|^2 +  g_{\alpha} |\psi_{\alpha}|^4   -2 u_2  (\psi_1^{\dagger}\psi_2 + \psi_2^{\dagger}\psi_1  )^2 +\cdots\nonumber
\end{eqnarray}
with $\alpha={1,2}$. The site index ($i,j$) on a square lattice are introduced. The link-variable $a_{ij}$ is for the $U(1)$ gauge potential, and its field strength, $f(i_p)$, is defined on a dual lattice with a plaquette index $i_p$. 
Several parameters including a gauge charge $e$ may access a variety of phases and their transitions. 
Microscopic information of the bosonic fields may be associated with doped spin liquid physics and naturally connected with fractionalization of valence-bond-solid orders as in the recent work  \cite{SachdevTop} (See SI).  
In this work we focus on a phenomenological model, leaving microscopic discussion to future works.
A $Z_2$ symmetry breaking transition may be realized by tuning $u_2$, and its low energy theory becomes  
\begin{eqnarray}
\mathcal{S}_{eff}& =& \int_x \big[ \frac{(\partial_{\mu} \rho)^2}{2\rho} + \frac{\rho}{2} (\partial_{\mu} \phi_+ -2 a_{\mu})^2  + \frac{r}{2} \rho + \frac{g}{4}\rho^2 +\frac{f^2}{2 e^2}+\cdots \big] \nonumber \\
&+&  \int_x  \big[ \frac{\rho}{2} (\partial_{\mu} \phi_-)^2 - 2 u_2 \rho \cos(2\phi_-) \big],
\end{eqnarray} 
with $\phi_{\pm} = \phi_1 \pm \phi_2$. For simplicity, we consider the case $(r_1, g_1) \simeq (r_2, g_2)$  giving $\rho = \rho_1 \simeq \rho_2 \neq 0$ whose differences may be treated perturbatively.  
The second line of the action is similar to the $Z_2$ clock model with $\phi_-$. 
Notice that $\phi_-$ is gauge neutral while $\phi_+$ has gauge charge two coupled to a U(1) gauge field. 
Accordingly, vortex configurations of $\phi_-$ is free of gauge flux, but ones with $\phi_+$ is attached to gauge flux with the flux-quantization rule, $\varPhi_{flux} = \oint a_{\mu} d x^{\mu}= (2\pi n_+^v)/2$.

We stress that vortex configurations of the two variables $(\phi_-, \phi_+)$ are not independent. 
The vortex numbers are defined by $n_{\pm}^v = n^v_{1}\pm n^v_2$ with ${2\pi} n_{1,2}^{v} =  \oint dx^{\mu}   \partial_{\mu}  \phi_{1,2}  $, and the integer variables  ($n^v_{1,2}$) indicate that the parity of $n^v_{+}$ is the same as  the one of $n^v_-$. 
The $2\pi$ vortex of $\phi_-$ indicates an odd integer $n_+^v$. By controlling a Hamiltonian of the  $(\phi_+, f)$ sector, we may suppress  the $2\pi$ vortex of $\phi_-$.
One obvious way for the suppression is to pay energy penalty to field strength by tuning the field strength energy terms,  $\frac{f^2}{2 e^2} + \frac{f(i_p)^4}{4 \Lambda}+\cdots $. 
For example, taking the limit of $\Lambda \rightarrow 0$, the energy penalty enforces  $n_+\rightarrow 0$ with an even number of $n_-$
Thus, the $2\pi$ vortex is suppressed energetically. 

Alternatively, one can use non-local interactions of the $(\phi_+, f)$ sector to suppress the $2\pi$ vortex of $\phi_-$.
Let us consider a lattice model, 
\begin{eqnarray}
S_{\vartheta} &=& -\frac{2\kappa}{\pi} \sum_{i,\mu,\alpha} \cos(\Delta_{\mu} \phi_\alpha - a_{\mu}) + \sum_{i_p} \frac{f(i_p)^2}{2 e^2} + i \frac{\vartheta}{2\pi} f(i_p) \nonumber
\end{eqnarray}
, where the amplitude $\rho$ in $S_{eff}$ is treated to be fixed with a coupling constant $\kappa$. 
The partition function $Z_{\vartheta} = \int_{[a] [\phi_1] [\phi_2]} e^{-S_{\vartheta}}$ may be analytically solved  treating the $u_2$ term perturbatively. 
By using the standard Villain approximation, 
we find  
\begin{eqnarray}
Z_{\vartheta} 
&=&  \sum_{[m_1],[m_2]}  \int_{[a]} {\rm exp} \Big[- \sum_{r,\mu} \frac{(\Delta_{\mu} m_1)^2+(\Delta_{\mu} m_2)^2}{4 \kappa/\pi} \Big]\nonumber \\
&\times&  {\rm exp} \Big[ -\sum \big(\frac{f^2}{2 e^2}   \big)  +i f(m_1+m_2+   \frac{\vartheta}{2\pi}) \Big]   \nonumber
\end{eqnarray} 
where the integer variables $m_{1,2}$ are defined on a dual lattice. 
It is obvious that the role of the  non-local term ($\vartheta=\pi$) is to shift the integer variable $m_1+m_2$ by a half-integer. Microscopically, this term is related to a non-local interaction term such as a background object associated with a half-flux (see SI). 
The Poisson summation formula allows to extract the $\vartheta$ contribution,
\begin{eqnarray}
Z_{\vartheta}  
&=&  \sum_{[n^v_1],[n^v_2]} \prod_{i_p}\Big( \cos(\frac{\vartheta}{2} (n^v_1+n^v_2))\Big) F[n^v_1,n^v_2]. 
\end{eqnarray}
The explicit form of $F[n^v_1,n^v_2]$ is presented in SI.  
Remarkable parity effects of $n_+^v = n_1^v+n_2^v$ are manifest. 
Vortex configurations with an odd $n_+^v$ is completely suppressed demonstrating the suppression of $2\pi$ vortex of $\phi_-$. 
Remark that the action with the $\vartheta$ term can be related to  quantum spin-chain in the easy-plane limit \cite{Haldane}. 

The suppression of the $2\pi$ vortex gives the dual theory,  
\begin{eqnarray}
\mathcal{S}_{dual}^{{\rm ICM}} = \int_x   \frac{(\partial_{\mu} \tilde{\phi})^2}{2 \pi \kappa}  - u_{4v} \cos(4\tilde{\phi}) - u_2 \cos(2\phi),
\end{eqnarray}  
where we drop the subscript $(-)$ hereafter.  
The $4\pi$ vortex term is manifest with the cosine term, $\cos(4\tilde{\phi})$.  
Interestingly, the form of the cosine terms is similar to the one of $\mathcal{S}_{Z_4,dual}$ with inverted-structures. 
Thus, we dub $S_{dual}^{{\rm ICM}} $ the inverted clock model (ICM), and it is easy to construct a mapping between ($\phi, \tilde{\phi}, \kappa$) and ($\Phi, \tilde{\Phi}, K$) indicating a critical theory of the ICM model has $c=1$. 
 
Striking properties of the universality class of ICM (ICM class) can be investigated by modifying analysis of the clock models \cite{Jose, Kadanoff}.  
Scaling dimensions of the cosine terms are $[\cos(4\tilde{\phi})] = 4\kappa$ and $[\cos(2 \phi)] = 1/\kappa$ near the Gaussian fixed point. 
The critical value of $\kappa$ is $\kappa_c =1/2$, where the two cosine terms are marginal. Away from $\kappa_c$, either one of the cosine terms becomes relevant. 
The scaling dimension of an order parameter is $[ e^{i\phi}] =1/2$, which gives the anomalous dimension and external field exponent, $\eta_{ICM}=1$ and $\delta_{ICM}=3$ in sharp contrasts to ones of the Ising class, $\eta_{Ising}=1/4$ and $\delta_{Ising}=15$. These are summarized in Table I.

 \begin{table}
\begin{tabular}{ ||c|| c || c || c||  } 
\hline
\,\,\,\, Univ. Class \,\,\,\,&\,\, $[\Delta]$ \,\,&\,\,\,\, $\eta$ \,\,\,\,&\,\,\,\, $\delta$ \,\,\,\,\\
\hline \hline
\,\, Ising (2D) \,\,& $1/8$ & $1/4$ & $15$ \\
\hline
\,\, MFT (4D) \,\,& $1$ & $0$ & $3$ \\
\hline \hline
\,\, ICM (2D) \,\,& $1/2$ & $1$ & $3$ \\
\hline
\end{tabular}
\caption{Universality class with a $Z_2$ symmetry, and their critical exponents. The order parameter scaling dimension $[\Delta]$, anomalous dimension $\eta$, and external field dependence exponent $\delta$ are presented.  MFT is for mean field theory, and ICM is for the universality class of the inverted clock model.   
}
\end{table}

The renomalization group (RG) equations are obtained by modifying the RG analysis of the $Z_4$ clock model \cite{Jose, KT, Kadanoff, Ginsparg} (see SI, also), 
\begin{eqnarray}
\frac{d}{dl} H_1= H_2 H_3, \quad \frac{d}{dl} H_2= H_1 H_3, \quad \frac{d}{dl} H_3= H_1 H_2 \nonumber
\end{eqnarray}
upto the leading order with the renormalization parameter $l$ with $H_1=4(\kappa-\frac{1}{2})$, $H_2 = 2(u_{4v}+u_2)$, and $H_3= 2(u_{4v}-u_2)$. 
There are critical lines ($H_{i}=H_j=0$ for $i \neq j$) with the multi-critical point at the origin ($H_{1,2,3}=0$). Remarkably, critical exponents  are {non-universal}, so critical properties depend on how to approach the critical lines. Near the origin, three different behaviors  of the correlation length are obtained by analysing the RG equation with the different limits; 
\begin{eqnarray}
1)\,\, && \xi_{KT} \sim {\rm exp}({\frac{a}{\sqrt{|t|}}})\quad \textnormal{in the limit of } H_i=H_j \,\, \textnormal{and } \,\,   i \neq j \nonumber \\
2)\,\, && \xi_{SU(2)} \sim {\rm exp}({\frac{a}{{|t|}}})\quad \textnormal{in the limit of } H_1=H_2=H_3 \nonumber \\
3)\,\, && {\xi}_{power} \sim  |t|^{-1/H_2}\quad \textnormal{for} \,\, H_2\neq0 \,\, \textnormal{and } H_2 \ll 1. \nonumber
\end{eqnarray}
 The dimensionless temperature $t \equiv \frac{T}{T_c}-1$ and a non-univeral positive constant ($a$) are introduced. 
 The first and second limits have the same RG equations of the KT transition and SU(2) spin chain. 
The thrid one is obtained by $\frac{d}{dl} (H_1 \pm H_3) = \pm H_2 (H_1 \pm H_3)$ with $H_2 >0$ giving $\nu={1/H_2}$. 
The scaling analysis further gives the order parameter onset behaviors, $\Delta \equiv \langle e^{i \phi} \rangle \sim \xi^{-1/2}$. Each correlation length dependence gives different onsets of the order parameter. 
For example, with $\xi_{KT}$, the order parameter onset is exponential, $\Delta \sim e^{-c/\sqrt{|t|}}$ for $T < T_c$. 
Remarkably, the order parameter onset of the ICM class may be super-linear, which is distinctly different from the sub-linear onset of the Ising-class with $\beta=1/{8}$.

Notice that the $2\pi$ vortex suppression endows  $\mathcal{S}_{dual}^{{\rm ICM}}$ with the inverted structure of $\mathcal{S}_{Z_4, dual}$.  If  $4\pi$ vortex configurations are further suppressed, then the ICM class may have an inverted structure of the $Z_6$ clock model. 
In analogy with the $Z_6$ clock model, the symmetry breaking transition is in the KT universality class with the correlation length, $\xi_{KT}$ \cite{Jose, KT}. One plausible mechanism of $4\pi$ vortex suppression is to incorporate a non-trivial quantum number inside vortex cores as shown in  \cite{Levin, Grover, qbt} and impose corresponding symmetry.  Doped quantum spin liquids, where fermionic excitations naturally appear, may naturally host mechanisms of  the suppression. 

We further discuss implication of a super-linear onset of an order parameter. 
In 2D, powerful structures of CFTs guarantee that the Landau theory with a $Z_2$ symmetry is in the Ising class with $\beta=1/8$, and thus a super-linear onset is incompatible with a $Z_2$ symmetry under the Landau paradigm. 
In the ICM class, we resolve the incompatibility by going beyond the Landau paradigm. 
The suppression of the $2\pi$ vortex induces the CFT with $c=1$. 
An alternative way is to enlarge a symmetry, say from $Z_2$ to $Z_4$, and then, a super-linear onset of an order parameter may be realized, the Ashkin-Teller CFT with $c=1$, even under the Landau paradigm.     
Therefore, observation of a super-linear onset in experiments indicates that either a broken symmetry is bigger than $Z_2$ or exotic transition beyond the Landau paradigm appears.

 %
%%%%%%%%%%%%%%%%%%%%%%%%%%%%%%%%%%%%%%%%%%%%%%%%%%%%%%%%%%%%%%%%%%%%%%%%%%%%%%
%%%%%%%%%%%%%%%%%%%%%%%%%%%%%%%%%%%%%%%%%%%%%%%%%%%%%%%%%%%%%%%%%%%%%%%%%%%%%%
\begin{figure}
\includegraphics[width=3.2in]{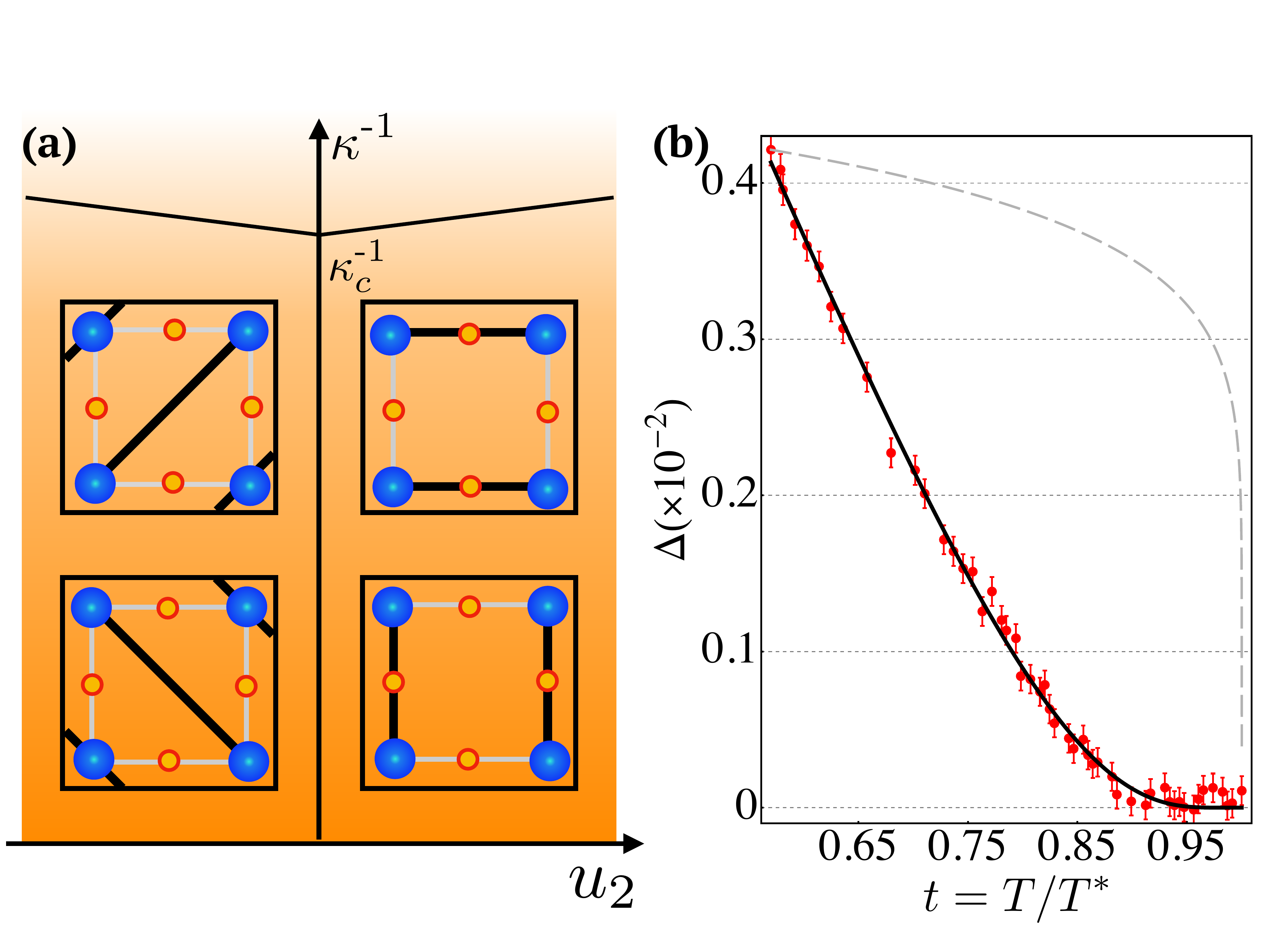}
\caption{(a) Schematic phase diagram of the inverted clock model.  The inverse temperature parameter $\kappa$ tunes $Z_2$ symmetry breaking transitions.  The sign of $u_2$ determines different patterns. For example, nematic transitions on a square lattice ($C_4 \rightarrow C_2$) have either bond or diagonal nematicity, which is classified by a reflection symmetry (say, a bond direction). The insets are to illustrate different patterns from the sign of $u_2$.  The filled circles are for sites of a square lattice and  the black lines are to distinguish bond ($u_2 >0$) and diagonal nematicity ($u_2 <0$).   (b) Order parameter onset of the ICM class, $\Delta (T) = A e^{-\frac{B}{\sqrt{1-T/T^*}}}$ (the KT type onset). Data (red dots) fitting of the magneto torque experiment \cite{nem3} is presented. The dashed line is a typical onset of the Ising class.}
\end{figure}
%%%%%%%%%%%%%%%%%%%%%%%%%%%%%%%%%%%%%%%%%%%%%%%%%%%%%%%%%%%%%%%%%%%%%%%%%%%%%%
%%%%%%%%%%%%%%%%%%%%%%%%%%%%%%%%%%%%%%%%%%%%%%%%%%%%%%%%%%%%%%%%%%%%%%%%%%%%%%

Let us apply our results to recent experiments in cuprates which report super-linear onsets of a nematic order around pseudogap temperatures  \cite{nem1, nem2, nem3}.
We perform data-fitting analysis of the magnetic torque data of \cite{nem3} with the KT-type onset, 
$\Delta (T) = A e^{-\frac{B}{\sqrt{1-T/T^*}}} $
and positive constants $A,B$ as illustrated in  Fig. 1. 
External symmetry breaking effects from the chain structure are subtracted as in the analysis of the experiments (See SI, for more details). 
Based on almost perfect match over wide range of temperatures, we propose that   { phenomena around pseudogap temperatures may be described by a continuous phase transition of  a CFT with $c=1$}. 

We provide three phenomenological scenarios for the $c=1$ transition. 
First, the transition is under the Landau paradigm with a bigger symmetry group, say $Z_4$. 
An Ising-nematic order cannot describe the transition, but the order parameters such as loop-current \cite{Varma} or d-density wave \cite{sudip} orders may be plausible. However, the onset of such order parameters requires additional experimental signatures. Namely, time-reversal (translational) symmetry breaking must be broken by loop-current order (d-density wave), but signals of the broken symmetries are under debates in spite of manifest signatures of the broken rotational symmetry.  
Second, the transition is beyond the Landau paradigm and only $Z_2$ symmetry is broken as in the ICM class.
The presence of the enigmatic strange metal and violation of the Luttinger theorem in under-doped cuprates \cite{Badoux} seems naturally connected to this scenario.  This scenario may also be related to deconfinement of doped quantum spin-liquids and the recent work about topological orders \cite{Lee, Sachdev, SachdevTop} (see SI). 
The final one is that the transition is beyond the Landau paradigm and the broken symmetry is $Z_4$. The universality class may be described by another Sine-Gordon model with the self-dual cosine terms with $\cos(4\phi)$ and $\cos(4\tilde{\phi})$. Here, the order parameter onset is also the KT type, and there is additional KT phase transition at high temperature. Note that all of symmetry breaking transitions of the three scenarios are associated with the CFT with $c=1$.

In conclusion, we present mechanisms of $Z_2$ symmetry breaking beyond the Landau paradigm by employing fractionalizion of discrete order parameters.
Characteristic critical exponents and correlation length behaviors are obtained. 
We find that the onset of an order parameter may be {super-linear} in contrast to the sub-linear onset of the Ising class. 
Our results of the non sub-linear onset  are applied to recent experiments in cuprates.  
Further theoretical studies including disorder effects and numerical analysis of microscopic lattice models in terms of valence bond solid or loop-current are highly desired.

 {\it Acknowledgement :} 
 We thank E. Berg, A. Furusaki, J. H. Han, S. Lee, Y. Matsuda, C. Mudry, N. Nagaosa, T. Shibauchi, A. Tanaka, and  K. Totsuka for invaluable discussion and comments. We are grateful to Y. Matsuda for providing the magneto-torque data. EGM is grateful to A. Furusaki, S. Lee, Y. Motome, and M. Sato for their hospitalities during the visits to RIKEN, KIAS, University of Tokyo, and YITP.  
This work was supported by the POSCO Science Fellowship of POSCO TJ Park Foundation and NRF of Korea under Grant No. 2017R1C1B2009176.

\newpage
\appendix

\section{Representations with fractionalized degrees of freedom}

Let us consider representations with fractionalized degrees of freedom,
\begin{eqnarray}
B_{1} = \rho_{s_{1}} \rho_1 \cos(2\phi_1 - 2\phi_{s_{1}}), \quad B_{2} = \rho_{s_{2}} \rho_2 \sin(2\phi_2 - 2\phi_{s_{2}}),  \nonumber
\end{eqnarray}
with the transformations of the $\pi/2$ rotation ($\mathcal{R}_{\pi/2}$) and the reflection ($\sigma_y : y \rightarrow -y$),
\begin{eqnarray}
(\phi_1, \phi_2,  \phi_{s_{1,2}}) &\rightarrow& (\phi_1 + \frac{\pi}{2}, \phi_2 - \frac{\pi}{2}, \phi_{s_{1,2}}) \quad \quad (\mathcal{R}_{\pi/2}) \nonumber \\
(\phi_1, \phi_2,  \phi_{s_{1,2}}) &\rightarrow& (-\phi_1 , -\phi_2, -\phi_{s_{1,2}}) \quad \quad \quad \quad (\sigma_y). \nonumber 
\end{eqnarray}
Then, the condensation of the spectator fields $\phi_{s_{1,2}}$ makes the representation conventional. 
Other than the conventional order parameters, one can have emergent order parameters, 
\begin{eqnarray}
b_{1} = \sqrt{\rho_1 \rho_2} \cos(\phi_1 - \phi_2), \quad b_{2} =\sqrt{\rho_1 \rho_2} \sin(\phi_1 - \phi_2), \nonumber
\end{eqnarray}
which transform precisely as $(B_{1}, B_{2})$.

\section{Lattice model analysis}
In this section, we discuss how to realize the lattice model with the non-local interaction in the main-text,  
\begin{eqnarray}
S_{\vartheta} &=& -\frac{2\kappa}{\pi} \sum_{i,\mu,k} \cos(\Delta_{\mu} \phi_k - a_{\mu}) + \sum_{i_p} \frac{f(i_p)^2}{2 e^2} + i \frac{\vartheta}{2\pi} f(i_p). \nonumber
\end{eqnarray}
The standard notation for lattice derivation is used and the gauge potential $a_{\mu}$ is defined on links. The field strength $f$ is defined on the dual lattice.
The presence of the unusual $\vartheta$ term may be understood by introducing high energy degrees of freedom with a half of the gauge charge of $\phi_{1,2,}$. Let us consider an action, 
\begin{eqnarray}
S_{UV } = S_{\vartheta=0} -\frac{t}{\pi} \sum_{i,\mu} \cos(\Delta_{\mu} \Theta - \frac{1}{2} a_{\mu}). \nonumber
\end{eqnarray}
The Villain approximation gives
\begin{eqnarray}
e^{t  \cos(\Delta_{\mu} \Theta - \frac{t}{2 \pi} a_{\mu})} \rightarrow \sum_{M_{\mu}} e^{-\frac{1}{2} (\Delta_{\mu} \Theta - \frac{1}{2} a_{\mu} - 2\pi M_{\mu})^2 } \nonumber
\end{eqnarray}
and the Poisson formula gives 
\begin{eqnarray}
 \sum_{J_{\mu}} e^{-\frac{\pi}{2 t} J_{\mu}^2 - i J_{\mu} \Delta_{\mu} \Theta + i \frac{1}{2} J_{\mu} a_{\mu}  }. \nonumber
\end{eqnarray} 
The integer variables $J_{\mu}, M_{\mu}$ on links are introduced. 
The angle variable integration endows the delta function, which may be replaced by $J_{\mu} = \epsilon_{\mu \nu} \Delta_{\nu} N$ with an integer variable $N$ on plaquette sites. 
The action becomes 
\begin{eqnarray}
S_{eff} = S_{\vartheta=0} + \sum_{i_p} i  \frac{ N}{2} f(i_p) + \frac{\pi}{2 t} (\Delta_{\mu} N)^2 \nonumber
\end{eqnarray}
with a plaquette index $i_p$.
It is obvious that in the limit of $t \rightarrow \infty$ the fluctuation term vanishes and $S_{eff}$ with the $N=1$ background becomes equivalent to  $S_{\vartheta=\pi}$.

With the lattice model $S_{\vartheta}$, one can analytically calculate the partition function,
\begin{eqnarray}
Z_{\vartheta} = \int_{[a] [\phi_1] [\phi_2]} e^{-S_{\vartheta}}.
\end{eqnarray}
Again, using the Villain approximation, the Poisson summation formula, and the $\phi_{1,2}$ integrations, we find 
\begin{eqnarray}
Z_{\vartheta} &=& \int_{[a]}\sum_{l_{1\mu}=-\infty}^{l_{1\mu}=+\infty} {\rm exp} \Big[- \sum_{r,\mu} \big(\frac{l_{1\mu}^2+l_{2\mu}^2}{4 \kappa/ \pi} + i ( l_{1\mu} + l_{2\mu} )a_{\mu} \big) \Big] \nonumber \\
&\times&  {\rm exp} \Big[ -\sum \big(\frac{f^2}{2 e^2} + i \frac{\vartheta}{2\pi} f \big) \Big] \prod_{r} \delta (\Delta_{\mu} l_{1\mu})  \delta (\Delta_{\mu} l_{2\mu}) \nonumber
\end{eqnarray}
The delta functions can be rewritten by introducing the number variables  $n_1,n_2$ on the dual lattice,  
\begin{eqnarray}
l_{1\mu} = \epsilon_{\mu \nu} \Delta_{\nu} n_1, \quad  l_{2\mu} = \epsilon_{\mu \nu} \Delta_{\nu}n_2. \nonumber
\end{eqnarray}
The gauge field integration gives the partition function,  
\begin{eqnarray}
Z_{\vartheta}
&=& \sum_{n_1,n_2}  {\rm exp} \Big[- \sum_{r,\mu} \frac{(\Delta_{\mu} n_1)^2+(\Delta_{\mu} n_2)^2}{4 \kappa/\pi} \Big] \nonumber\\
&&\times {\rm exp} \Big[ -\sum \frac{e^2}{2} (n_1+n_2+\frac{\vartheta}{2\pi})^2 \Big] \nonumber
\end{eqnarray}
Using the Poisson formula again and rescaling the fields $\phi_{1,2} \rightarrow 2\phi_{1,2}/\pi$, we have 
\begin{eqnarray}
Z_{\vartheta}
&=&  \int_{[\tilde{\phi}_1] [\tilde{\phi}_2]} \sum_{m_1,m_2} {\rm exp}\Big[- \sum_{r,\mu} \frac{(\Delta_{\mu} \tilde{\phi}_1)^2+(\Delta_{\mu} \tilde{\phi}_2)^2}{\pi \kappa} \Big] \nonumber\\
&\times& {\rm exp} \Big[i 4 (\tilde{\phi}_1 m_1 +\tilde{\phi}_2 m_2) +\frac{e^2}{2} (\frac{2(\tilde{\phi}_1+\tilde{\phi}_2)}{\pi}+\frac{\vartheta}{2\pi})^2 \Big]. \nonumber
\end{eqnarray}
Introducing $\phi_{\pm} = \phi_1 \pm \phi_2$ and shifting the variable, we have  
Shifting the variable $\phi_+ \rightarrow \phi_{+} -\frac{\alpha}{4}$, we obtain
\begin{eqnarray}
Z_{\vartheta} 
&=&  \sum_{m_1,m_2} e^{\sum_{r} -i  \frac{\vartheta}{2}(m_1+m_2)} F[m_1,m_2]  \nonumber\\
F[m_1,m_2] &=&\int_{[\tilde{\phi}_-] [\tilde{\phi}_+]}  {\rm exp}\prod_{r,\mu} \Big[- \frac{(\Delta_{\mu} \tilde{\phi}_+)^2+(\Delta_{\mu} \tilde{\phi}_-)^2}{2\pi \kappa} \Big] \nonumber\\
&\times& {\rm exp} \Big[i (2\tilde{\phi}_+ (m_1+m_2) + 2\tilde{\phi}_- ( m_1-m_2) \Big] \nonumber \\
&\times& {\rm exp} \Big[+\frac{e^2}{2} (\frac{2\tilde{\phi}_+}{\pi})^2 \big)\Big] \nonumber
\end{eqnarray}
It is obvious to see that $F[m_1,m_2]=F[-m_1,-m_2]$, so we have the final form, 
\begin{eqnarray}
Z_{\vartheta}  
&=&  \sum_{n^v_1,n^v_2} \prod_{i_p}\Big( \cos(\frac{\vartheta}{2} (n^v_1+n^v_2))\Big) F[n^v_1,n^v_2] \nonumber
\end{eqnarray}
The destructive interference with $\vartheta=\pi$ for an odd number of $n_+^v$ is obvious. 

The parity of $n^v_+$ is the same as one of $n_-^v$, and thus the minimal vortex of $\phi_-$ is not a $2\pi$ vortex but a $4\pi$ vortex. 
Moreover, the $\tilde{\phi}_+$ is short-ranged by the presence of $e^2$ term, so the low energy theory becomes
\begin{eqnarray}
\mathcal{S}_{\vartheta, dual} = \int_x   \frac{(\partial_{\mu} \tilde{\phi}_-)^2}{2 \pi \kappa}  - u_{4v} \cos(4\tilde{\phi}_-). 
\end{eqnarray} 
As usual in the derivation of the Sine-Gordon model, we put the fugacity term of the vortex, $u_{4v}$, by hand.   
Notice that the $\vartheta$ term naturally appears in $s=1/2$ spin chains, and our discussion on the exotic $Z_2$ transition with the inverted clock model may be directly applied to spin $s=1/2$ chain.

\section{Properties of the inverted clock model}
In this section, we explicitly present differences between the $Z_4$ clock model and the inverted clock model.
The $Z_4$ clock model is   
\begin{eqnarray}
\mathcal{S}_{Z_4} = \int_x \frac{{K}}{2\pi } (\partial_{\mu} {\Phi})^2 - u_{4} \cos(4{\Phi}), \nonumber
\end{eqnarray}
whose dual theory is 
\begin{eqnarray}
\mathcal{S}_{Z_4, dual} = \int_x \frac{ 1}{2\pi {K}}  (\partial_{\mu} \tilde{\Phi})^2 - u_{2v}  \cos(2\tilde{\Phi}) - u_{4} \cos(4{\Phi}). \nonumber
\end{eqnarray}
Around the Gaussian point, the scaling dimensions of the two cosine potential scaling are 
\begin{eqnarray}
[\cos(4\Phi)] = \frac{4}{K}, \quad [\cos(2 \tilde{\Phi})] = K, \nonumber
\end{eqnarray}
and at $K= K_c=2$, the two operators are marginal. Away from $K_c$, either one of the operators become relevant while the other one is irrelevant. 
Also, the scaling dimension of the $Z_4$ order parameter, $\Delta_{Z_4}  = \langle \cos(\Phi) \rangle$, is determined by  
\begin{eqnarray}
 [\cos(\Phi)] = \frac{1}{4 K}. \nonumber
\end{eqnarray}  
At $K_c$, the order parameter scaling dimension is $[\Delta_{Z_4}]=1/8$.

The inverted clock model with the $2\pi$ vortex suppression is  
\begin{eqnarray}
\mathcal{S}^{ICM}_{dual} = \int_x \frac{1}{2\pi \kappa } (\partial_{\mu} {\tilde{\phi}})^2 - u_{4v} \cos(4 {\tilde{\phi}}) - u_2 \cos(2 \phi). \nonumber
\end{eqnarray}
We drop the subscript $-$ for simplicity. 
Around the Gaussian point,  the scaling dimensions of the two cosine potential scaling are 
\begin{eqnarray}
[\cos(4\tilde{\phi})] = {4}{\kappa}, \quad [\cos(2 {\phi})] = \frac{1}{ \kappa}, \nonumber
\end{eqnarray}
and at $\kappa = \kappa_c =1/2$, the two operators are marginal. 
Also, the scaling dimension of the $Z_2$ order parameter of the ICM class, $\Delta_{Z_2}  = \langle \cos(\phi) \rangle$, is determined by  
\begin{eqnarray}
 [\cos(\phi)] = \frac{1}{4 \kappa}. \nonumber
\end{eqnarray}  
At $\kappa_c$, the order parameter scaling dimension is $[\Delta_{Z_2}]=1/2$.

The scaling dimension of the external field is determiend by 
\begin{eqnarray}
&&\int_x h_4 \cos(\Phi) \rightarrow [h_4] = 2 - [\cos(\Phi)] = \frac{15}{8}, \nonumber \\
&&\int_x h_2 \cos(\phi) \rightarrow [h_2] = 2 - [\cos(\phi)] = \frac{3}{2}. \nonumber
\end{eqnarray}
The scaling analysis gives 
\begin{eqnarray}
\Delta \propto (\xi^{-1})^{[\Delta]} \propto h^{\frac{[\Delta]}{[h]}} \rightarrow \delta = \frac{[h]}{[\Delta]},
\end{eqnarray}
which gives 
\begin{eqnarray}
\delta_{Z_4}= 15, \quad \delta_{Z_2, ICM} = 3. \nonumber
\end{eqnarray}
The suscpetibility can be obtained in a similar way, 
\begin{eqnarray}
\chi = \frac{\partial \Delta}{\partial h} \propto h^{\frac{[\Delta]-[h]}{[h]}} \propto (\xi^{-1})^{[\Delta]-[h]}. 
\end{eqnarray} 
Note that if $4\pi$ vortex is further suppressed, then similar analysis  to the $Z_N$ clock model with $N>4$ can be straightforwardly done.  
 
As mentioned in the main-text, further suppression of $4\pi$ vortex configurations makes the $Z_2$ symmetry breaking transition as in the KT class. There are three possible phases; a symmetry broken phase at low temperature, KT phase at intermediate phase, and disordered phase at high temperature. The two transitions are both described by the KT transitions \cite{Jose, KT}, and at the symmetry breaking transition temperature, the scaling dimensions of physical operators are determined by the condition, $[\cos(2\phi)]=2$, around the Gaussian point, giving
\begin{eqnarray}
[\Delta_{Z_2}] = \frac{1}{2}, \quad \delta=3.
\end{eqnarray}

Finally, it is possible that the system has a $Z_4$ symmetry with $2\pi$ vortex suppression. Then, the critical theory becomes
\begin{eqnarray}
\mathcal{S}_{Z_4,dual}^{ICM} = \int_x \frac{(\partial \tilde{\phi})^2}{2\pi \kappa} - u_{4v} \cos(4\tilde{\phi}) - u_4 \cos(4\phi). \nonumber
\end{eqnarray}
The scaling dimensions of the cosine terms are $[\cos(4\phi)]=\frac{4}{\kappa}$ and $[\cos(4\tilde{\phi})] = 4 \kappa$ around the Gaussian point. 
There are two critical points $\kappa_{c1}=1/2$ and $\kappa_{c2}=2$. At low temperature ($\kappa > \kappa_{c2}$), the $Z_4$ symmetry is broken and the criticality is described by the order parameter $\Delta_{Z_4} =\langle e^{i\phi} \rangle $ whose onset is the KT type with the scaling dimension, $[\Delta_{Z_4}]=1/8$. In the intermediate region $\kappa_{c1} < \kappa < \kappa_{c2}$, the KT phase appears.

\subsection{Renormalization group analysis}
The universality class of the ICM class can be investigated by applying the RG analysis of the clock model.
Near  the Gaussian point ($u_2=u_{4d} =0$), it is easy to obtain    
\begin{eqnarray}
\frac{d}{d l} \kappa &=& - u_{4d}^2 +4 \kappa^2 u_2^2 , \nonumber \\
\frac{d}{d l} u_{4d} &=& (2-4 \kappa) u_{4d}, \nonumber \\
 \frac{d}{d l} u_2 &=& (2-   \frac{1}{\kappa}) u_2. \nonumber 
\end{eqnarray}
Notice that the duality at the level of the partition functions ($\mathcal{Z}(\kappa, u_2, u_{4d}) = \mathcal{Z}_{dual}((4 \kappa)^{-1}, u_{4d}, u_2)$) are well established.  

Introducing new variables, 
\begin{eqnarray}
H_1 = 4(\kappa-\frac{1}{2}), \quad H_2 = 2 (u_2 +u_{4d}), \quad H_3=2 (u_2 -u_{4d}), \nonumber
\end{eqnarray}
the RG equations are simplified, 
\begin{eqnarray}
\frac{d}{dl} H_1= H_2 H_3, \quad \frac{d}{dl} H_2= H_1 H_3, \quad \frac{d}{dl} H_3= H_1 H_2, \nonumber
\end{eqnarray}
which is the same as the RG equations obtained by the work by Kadanoff \cite{Kadanoff}.

The critical lines are determined by the condition, $H_i =H_j =0$ and $i \neq j$. 
Near a fixed point with $H_2>0$ and $H_2 \ll 1$, the RG equation becomes
\begin{eqnarray}
\frac{d}{dl} (H_1 \pm H_3) = \pm H_2 ( H_1 \pm H_3). \nonumber
\end{eqnarray}
$H_1-H_3$ is irrelevant, and  $H_1 +H_3$ is relevant giving $\nu = 1/H_2$. 
Notice the scaling dimension can be very small in the limit, $H_2 \ll 1$.
It is easy to observe that the fine-tuned limit $H_1=H_2=H_3=H$ gives the RG equation, 
\begin{eqnarray}
\frac{d H}{d l} = H^2, \nonumber
\end{eqnarray}
and the correlation length becomes
\begin{eqnarray}
\xi \sim e^{\frac{c}{|t|}}. \nonumber
\end{eqnarray}

\section{Relations with microscopic models}
In this section, we discuss relations with microscopic models to realize the exotic $Z_2$ transitions with the ICM class. Especially, we show close relation to the previous work on topological order \cite{SachdevTop}. 
As mentioned in the main-text, one concrete way is to consider fractionalization of physical operators. 
Let us consider the Schwinger boson representation, which is useful to describe phases nearby conventional magnetic symmetry broken phases. 
We start with writing the anti-ferromagnetic component of spin operators, $\vec{S}(i) = (-1)^{i} \vec{n}(i) +\cdots $,
\begin{eqnarray}
\vec{n}(i) = \frac{1}{2} b^{\dagger}_{\alpha}(i) (\vec{\sigma})_{\alpha \beta} b_{\beta}(i),  \nonumber
\end{eqnarray}
The valence bond solid operator can be defined as 
\begin{eqnarray}
&&V_{x} (\vec{i}) \equiv \vec{S} ({\vec{i}}) \cdot \vec{S} ({\vec{i}+\hat{x}}) (-1)^{i_x}, \quad \vec{i}=(i_x, i_y) \nonumber \\
&&V_{y} (\vec{i}) \equiv \vec{S} ({\vec{i}}) \cdot \vec{S} ({\vec{i}+\hat{y}}) (-1)^{i_y}.\nonumber 
\end{eqnarray}
Rewriting the dimer operators in terms of the Schwinger bosons, one can show 
\begin{eqnarray}
V_{x} \propto (-1)^{i_x} Q_x^{\dagger} Q_x, \quad V_{y} \propto (-1)^{i_y} Q_y^{\dagger} Q_y, \nonumber 
\end{eqnarray}
with 
\begin{eqnarray}
Q_{x}(i) &=& \varepsilon_{\alpha \beta}  b_{i,\alpha}b_{i+x,\beta},  \quad 
Q_{y}(i) = \varepsilon_{\alpha \beta}  b_{i,\alpha}b_{i+y,\beta}  \nonumber. 
\end{eqnarray}
The operators ($Q_{x,y}$) have twice bigger gauge charge  than the Schwinger bosons. 
Then, one can interpret $Q_{x,y}$ as the VBS fractionalized operators. 
With some modifications, one can find a mapping between $Q_{x,y}$ and the fractionalized representation ($\psi_{1,2}$) in the main-text. The presence of the emergent order parameter is manifest, for example  $Q_x^{\dagger}Q_y+Q_y^{\dagger}Q_x$, which has the same symmetry transformation as $V_x V_y$.
More detailed discussion about the relations will be presented in other places.

\section{Data fitting}
\begin{figure}
\includegraphics[width=3.4in]{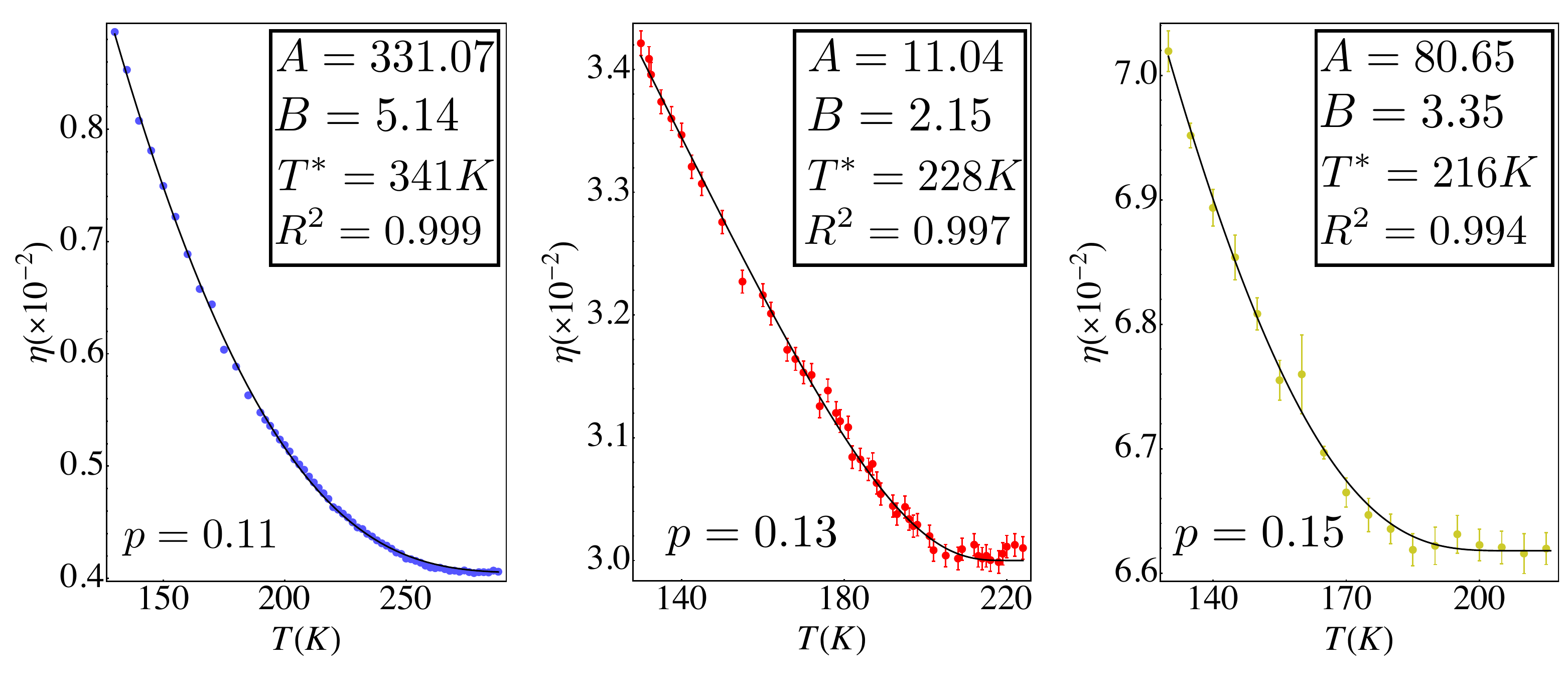}
\caption{ Fitting results of the torque experiments of \cite{nem3} with the KT onset, $\eta (T) = \eta(T > T^*) + A e^{-\frac{B}{\sqrt{1-T/T^*}}}$.
We read off the offset values $\eta(T>T^*)$ from the data, and the three parameters ($T^*$, $A$, $B$) are used. The statistical $R^2$ values are also presented. 
 } \label{fitting}
\end{figure}

The non sub-linear onset in the ICM class can be applied to the recent experiments in cuprates. 
As an example, we apply our results to the recent torque experiments in cuprates around the pseudo-gap temperatures. 
In the experiments, the materials (YBCO) have a spin chain structure along one of the bond directions, which plays a role of an external field. 
To take account of the spin chain effects, we modify the formula, 
\begin{eqnarray}
\eta (T) = \eta(T > T^*) + (\xi^{-1}(T))^{[\Delta]}.
\end{eqnarray}
Note that $T^*$ and $\eta(T>T^*)$ are experimentally observed. The off-set value $\eta(T>T^*)$ can be reasonably read off from the data, but the pseudo-gap temperature $T^*$ is not easy to be determined accurately because of the super-linear onsets. 
We first fit the data with two fitting parameters for the three types of the correlation functions, $\xi_{KT}$, $\xi_{SU(2)}$, and $\xi_{power}$ using the reported values of $\eta(T>T^*)$ and $T^*$. The $R^2$ in statistics is bigger than $0.99$ for all cases, and we find that $\xi_{KT}$ and $\xi_{power}$ show better fittings than $\xi_{SU(2)}$. Next, we also perform data fitting varying with $T^*$ because of uncertainty of $T^*$ in experiments, and we find that the fitting becomes better by setting the pseudo-gap temperatures higher than experimentally observed ones, and we present best results with the KT onsets in Fig.  \ref{fitting}.
We expect that systems without the spin-chain effects such as Hg compounds cuprates would be a better system to determine $T^*$ better. 
 
 %
%%%%%%%%%%%%%%%%%%%%%%%%%%%%%%%%%%%%%%%%%%%%%%%%%%%%%%%%%%%%%%%%%%%%%%%%%%%%%%
%%%%%%%%%%%%%%%%%%%%%%%%%%%%%%%%%%%%%%%%%%%%%%%%%%%%%%%%%%%%%%%%%%%%%%%%%%%%%%
%%%%%%%%%%%%%%%%%%%%%%%%%%%%%%%%%%%%%%%%%%%%%%%%%%%%%%%%%%%%%%%%%%%%%%%%%%%%%%
%%%%%%%%%%%%%%%%%%%%%%%%%%%%%%%%%%%%%%%%%%%%%%%%%%%%%%%%%%%%%%%%%%%%%%%%%%%%%%

  \end{document}